# Charge-transfer interfaces between metal and redox arylamine molecular films: As probed with anode interfacial engineering approach in single-layer organic diodes

J. C. Li,[a] S. C. Blackstock,[*] and G. J. Szulczewski[*]

Department of Chemistry and The Center for Materials for Information Technology, The University of Alabama, Tuscaloosa, AL 35487

We investigate the charge-transfer interfaces between metal and redox arylamine molecular films through studying the current–voltage characteristics of single-layer organic diodes with the aid of anode interfacial engineering method. The diode turn-on voltage is shown to be highly sensitive to the arylamine/metal charge-transfer interfaces and thus can serve as a probe in detecting such organic/metal interfaces. We show that the diode electrical performance could be tuned through engineering the arylamine/metal interfaces via controlling the factors of anode work function, arylamine substitute groups, and active layer surface morphology etc. The conduction mechanism of the diodes is shown to be injection limited, which could be well described with Richardson-Schottky thermonic emission model. Our work may provide some insight into the use of single-layer organic diode and interfacial engineering method to rapidly probe the organic/metal and even organic/organic charge-transfer interfaces.

[a]Present address: Department of Chemistry, The University of Chicago. E-mail: jcli@uchicago.edu
[*]Corresponding authors. E-mail: gjs@bama.ua.edu and blackstock@ua.edu

## Introduction

The field of organic electronics is developing at a rapid pace. So far, organic semiconductors have been using as active layers in various devices such as organic light emitting diodes, organic field effect transistors, and solar cells etc.[1-5] It is recognized that the organic/metal interfaces play a critical role in affecting the charge injection and device electrical performances.[6-8] The usual choice to examine metal/organic interfaces is to use ultra-high vacuum surface spectroscopies.[9-11] While such spectroscopy techniques permit the advantage of high resolution, it is often complicated and difficult to study the organic/metal interface in real device situations due to the limitation of film detection range. It is desirable to find some easy way to quantatively inspect the information of the organic/metal interfacial electronic structures in simple but real device conditions. To date, there are many advances in studying single-layer organic diodes and anode interfacial engineering of light-emitting diodes.[12-14] This may offer some new opportunities in investigating organic/metal interfaces, if we could combine their advantages together.

Arylamine derivatives have been widely studied and used as good hole transporting materials due to the easy oxidizablity of the nitrogen centers and the ability to transport positive charges via the radical cation species.[15] However, it is still not well understood how the cyano (-CN) substitutes affect the electrical properties of arylamine molecular films, arylamine/metal interfaces, and consequent device performance. Although -CN group is known to be strong electron acceptor, charge-transfer (CT) interactions are widely observed between metals and organic molecules with -CN substitutes.[16]

In the previous work, we showed that single-layer organic diode is a simple but effective way to characterize the electrical properties of arylamine molecular films and their devices.[17] We observed linear dependence between the molecular oxidation potential and diode turn-on voltage ($V_t$). However, extra-high $V_t$ is recently observed in diodes made from dendritic arylamines with –CN terminal groups, which does not obey such linear oxidation potential relationship. Cyclic voltammetry measurements indicate that the adding of –CN groups endows significant large redox-gradient to the arylamines, which may play a critical role in some potential applications especially in molecular-scale charge storage.[18] So it would be interesting to study the physical mechanism under such observation of ultra-high $V_t$ for developing redox arylamine-based organic electronics.

Here, we demonstrate how to use single-layer organic diode to quantitatively characterize the arylamine/metal CT interfaces. To the best of our knowledge, there is still no literature reported from the viewpoint of using single-layer organic diode as probe of organic/metal CT interfaces. Our strategy is simply based on inserting an interfacial modification layer between the anode and the arylamine film. The CT interface is then examined through analyzing the device I–V data with and without anode interfacial layer. The effect of arylamine film morphology on the CT interface and the device conduction mechanism are also briefly discussed.

**Experimental**

Figure 1 shows the schematic molecular structure of the arylamines used and the device configurations studied. Two dendritic redox arylamines are employed as both the hole transport materials and the interfacial modification layers. The synthesis, film growth, and electrochemical characterization of material 4AAPD have been reported.[19-20] The synthesis and electrochemical characterization of CN-4AAPD will be reported elsewhere. As seen, the CN-4AAPD has the same basic dimensions as that of the 4AAPD, but four of the anisyl groups in the periphery are replaced with cyano groups. The cyano groups are electron-withdrawing and cause the oxidation potential of the periphery amine groups to increase. The redox-gradient of CN-4AAPD is as high as 10.8 kcal/mol, which is almost two times as that of 4AAPD (5.1 kcal/mol).

The diodes have a structure of GaIn/arylamine(s)/Anode. They are fabricated by following consequent deposition of anode electrode and organic layer(s) onto clean glass slide without breaking the vacuum. A micrometer controlled liquid GaIn (as purchased from Aldrich) tip with diameter of about 100 micrometers is used as the cathode, which provides a non-damage soft−contact to the organic layer. The anodes (Au, Ag, and Cu) are all fresh vapor deposited 100-nm-thick film at the pressure of about $1.0 \times 10^{-6}$ Torr with deposition rate of 0.2 nm/s. The organic films are then deposited onto the anode with rate of 0.5 Å/s. The film thickness and deposition rate are measured by quartz crystal microbalance.

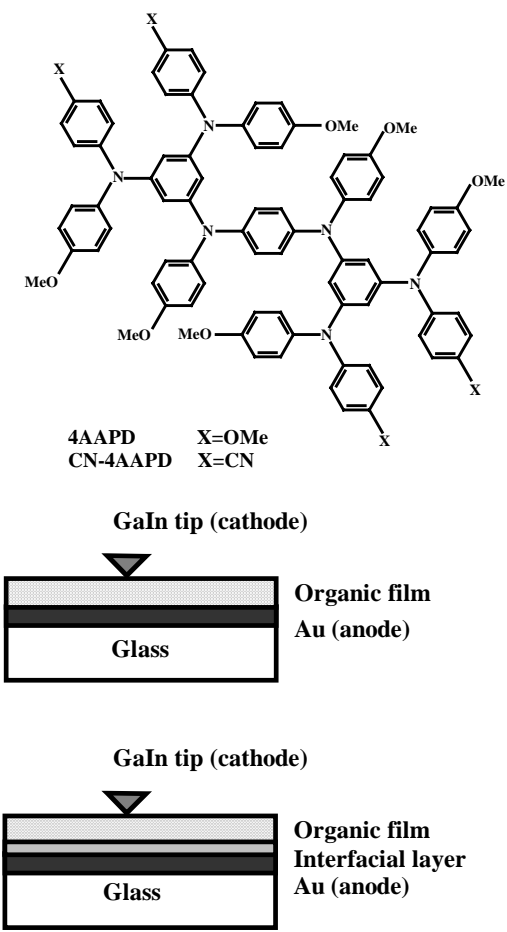

Fig. 1 Schematic molecular structure of the arylamines used and the device configurations studied.

Current−voltage (I–V) characteristics are measured from fresh samples by using a PAR-273 potentiostat (EG&G Instruments) with a 10 KΩ protecting resistor in ambient conditions. For accuracy, the data of $V_t$ is averaged from more than 25 measurements taken at different samples and various spots. The film morphology is analyzed with a Digital Instruments DI3100 atomic force microscopy (AFM) at tapping mode. Self-assembled monolayers (SAMs) of 1-Decanethiol are formed onto to fresh gold surface from 1 mM solution in pure ethanol for 24 hrs, then thoroughly rinsed with EtOH, and blow dry with nitrogen gas. We choose 1-Decanethiol because it is easy to get high quality monolayers on Au surface owing to its optimized alkane chain length.

**Results and discussion**

Figure 2 shows the typical I–V curves of the 4AAPD and CN-4AAPD diodes with a series of active layer thickness. In forward bias, the current exhibits nonlinear behavior after the applied voltage exceeding a threshold point, i.e., turn-on voltage $V_t$. Below $V_t$ and in reverse bias (even to our highest accessible -10 V), the samples only show leakage behavior with current value of about 0.1 nA. The thickness dependence of $V_t$ suggests that the current is electrical field mediated (as further discussed). Note that $V_t$ of CN-4AAPD diode is extraordinarily higher than that of the corresponding 4AAPD devices, although the oxidation potential of CN-4AAPD is only 0.16 V higher than that of 4AAPD ($E^o$ = 0.49 V vs. SCE). For example, $V_t$ of the 150 nm CN-4AAPD diode is almost 4 V higher than that of the 4AAPD device. Even for the 50 nm diodes, the difference is still about 1 V. This suggests that there may exist some large difference in the device electronic structure between 4AAPD and CN-4AAPD diodes.

The only structure difference between the two arylamines is the presence of four –CN groups in CN-4AAPD. Thus one possible reason is that the performance of CN-4AAPD diodes may be affected by a CT interaction between the arylamine and the electrodes, as indicated by the drastically increased $V_t$. To see if this is the case, we plot the current (I) versus effective electrical field ($E_{eff}$), with $E_{eff} = (V-V_t)/d$ (data not shown). Here, V and d denote the voltage drop on the sample and the active layer thickness, respectively. For 4AAPD diodes, the I–$E_{eff}$ plots collapse to one curve, indicating that the current is electrical field dependent. However, for CN-4AAPD diodes, the I–$E_{eff}$ plot of the 50 nm device is obviously apart from the collapsed curves of the 100 and 150 nm diodes. The current of 50 nm CN-4AAPD device is about 8 times lower than that of the others. The CT interaction is known to occur normally within a narrow film range (about 10 nm), which is therefore more dominant in device with thinner CN-4AAPD layer than that with thicker layer. This explains very well why the I–$E_{eff}$ curve of the thinner CN-4AAPD diode is different to that of the thicker devices.

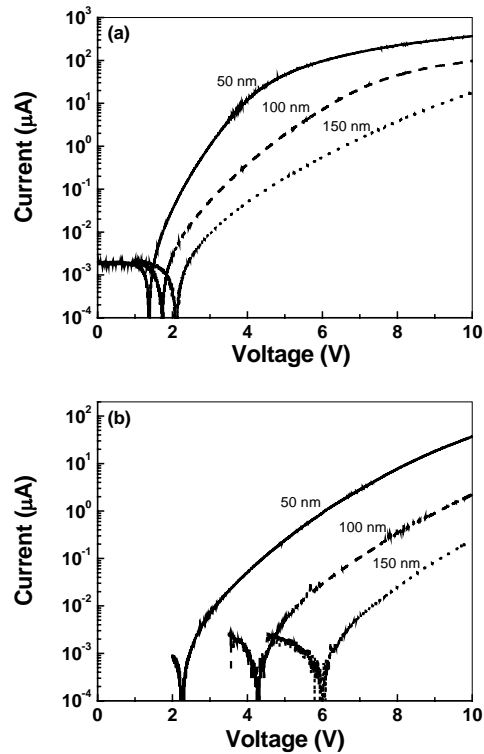

Fig. 2 Typical I-V curves of (a) GaIn/4AAPD/Au and (b) GaIn/CN-4AAPD/Au diodes with different active layer thickness.

Control experimental results confirm that there exists CT interaction at the CN-4AAPD/anode interface. Figure 3 shows the representative I–V curves of the 4AAPD and CN-4AAPD diodes with different anode modification layers. To simplify the comparison, the two arylamines are used as the interfacial layer for each other. When CN-4AAPD is used as the anode modification layer, $V_t$ of 4AAPD diodes is increased from about 1.4 to 2.8 V as the CN-4AAPD interfacial layer increases from 0 to 10nm, respectively. However, when 4AAPD is used as the anode interfacial layer in CN-4AAPD diodes, $V_t$ is reduced from 2.5 to 1.3 V as the 4AAPD interlayer increases from 0 to 5 nm, respectively. The difference between the highest and the lowest $V_t$ is more than 1 V. This may be interpreted that the charge injection barrier is changed with introducing an anode modification layer into the diodes. The CN-4AAPD and 4AAPD interlayer can effectively enhance and impede the CT interface, i.e., increase and decrease the hole injection barrier, between the anode and the arylamine molecular film, respectively. Fig.3(c) shows that the average thickness of 4AAPD interlayer is about 5-nm in order to largely block the CT interface in CN-4AAPD device. In contrary, one would need at least 10-nm-thick CN-4AAPD film as the interlayer to largely enhance the CT interface in 4AAPD devices. Moreover, these observations suggest that the CT interface can be effectively tuned with simply inserting a thin interfacial modification layer between the arylamine and anode.

The arylamine/anode CT interface can be obviously affected by alkanethiol SAMs adsorbed on the surface of Au anode. As shown in Fig. 4, both the 4AAPD and CN-4AAPD diodes with 1-Decanethiol SAMs interlayer would have a higher $V_t$ than that without such layer. For the 50 nm 4AAPD diodes, the device $V_t$ is increased about 0.5 V as compared with the corresponding diodes without SAMs interlayer. Surprisingly, $V_t$ of the CN-4AAPD devices is increased more than 2 V. As we know, alkanethiol monolayers may influence the work function of electrodes to a large extent.[11,21,22] Here, the 1-Decanethiol SAMs will decrease the Au anode work function. Consequently, the corresponding hole-injection barrier and further $V_t$ are increased for both 4AAPD and CN-4AAPD diodes. The slope of Fig. 4 (c) is just indicative of the threshold electrical field to turn-on the device, while the y-intercept indicates the build-in potential between the anode and the cathode. The bigger the slope, the higher the electrical field needed to turn-on the diode. The CN-4AAPD devices have a bigger slope than that of 4AAPD diodes. Note that the y-intercept value for CN-4AAPD on SAMs-Au is evidently higher than that of the others. We propose that, for GaIn/CN-4AAPD/SAMs/Au diodes, there may exist two CT interfaces: one is between 1-Decanethiol and Au, the other is at CN-4AAPD/SAMs-Au. The neat effect of the later interface equals to introduce an extra hole injection barrier

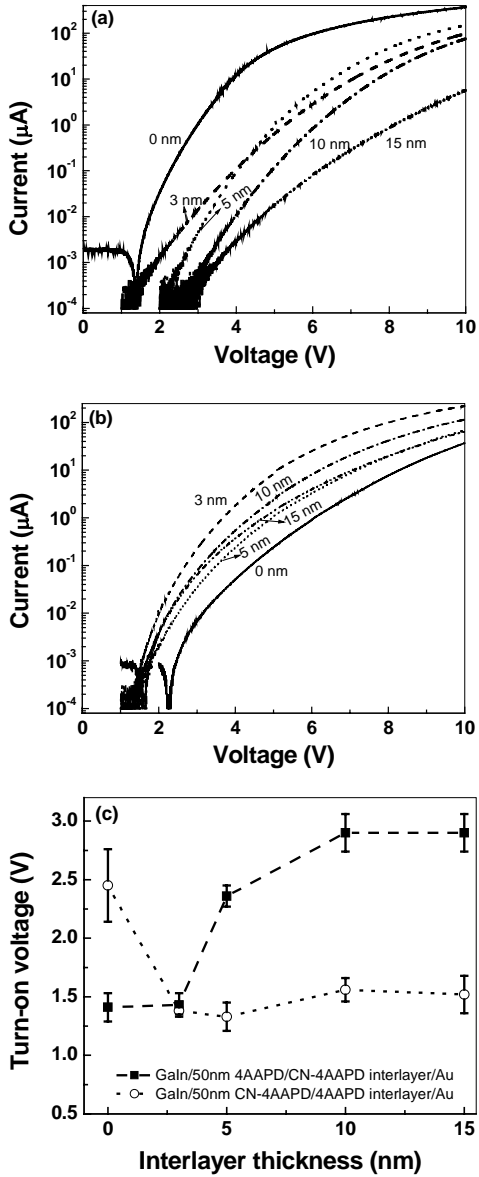

Fig. 3 Effects of anode modification layer on the charge-transfer interface. I–V curves of (a) GaIn/4AAPD (50 nm)/CN-4AAPD interlayer/Au, and (b) GaIn/CN-4AAPD (50 nm)/ 4AAPD interlayer/Au diodes with various interlayer thickness. (c) Plot of average turn-on voltage versus the interlayer thickness.

(or traces) to the diode. However, this interface should not be induced by the electron transfer from 1-Decanethiol to CN-4AAPD because of the wide energy gap of alkanethiol (~7 eV). It is reasonable to explain that the electron transfer is from the Au surface to the CN-4AAPD because the Decanethiol monolayers (less than 1.5 nm) cannot block such charge transfer.[23]

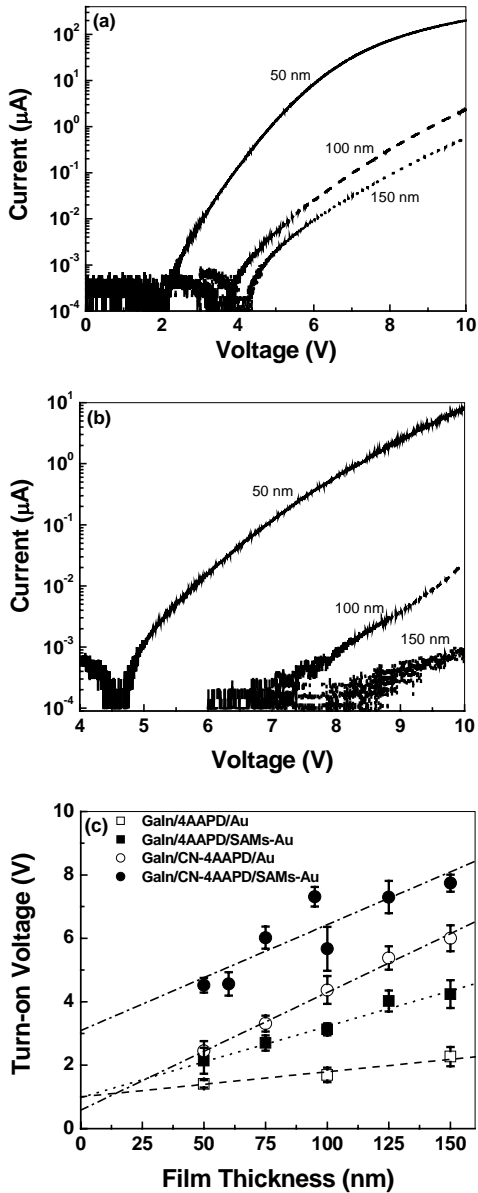

Fig. 4 Effect of 1-Decanethiol monolayers on the charge-transfer interface. Typical I-V curves of (a) GaIn/4AAPD/SAMs-Au, and (b) GaIn/CN-4AAPD/SAMs-Au diodes. (c) Thickness dependence of $V_t$ for the 4AAPD and CN-4AAPD diodes fabricated on Au and SAMs-Au, respectively.

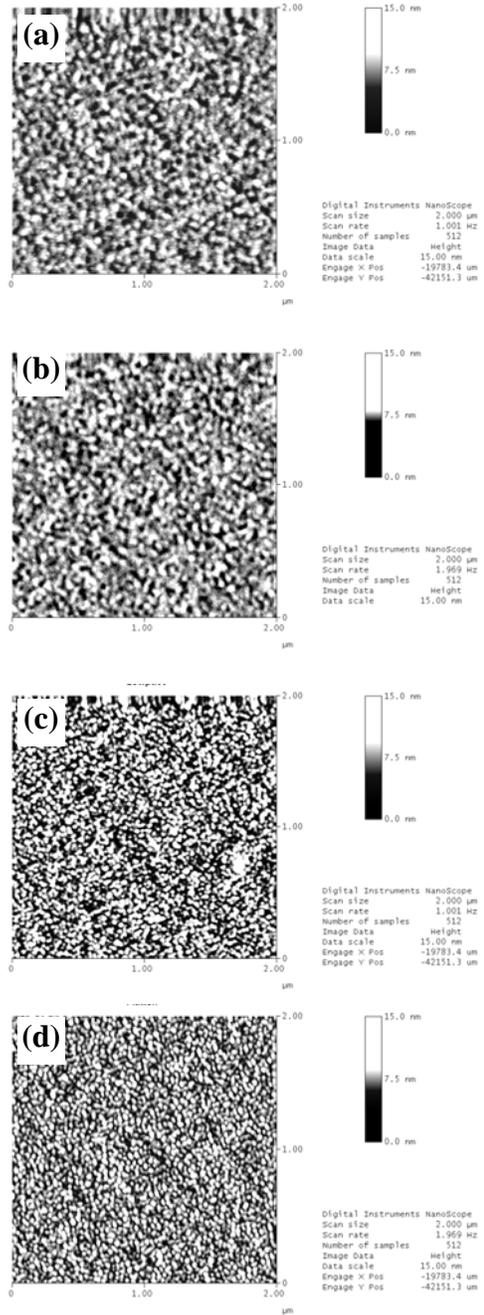

Fig. 5 Tapping mode AFM images of the 50 nm films of 4AAPD (a-b) and CN-4AAPD (c-d) deposited on Au and SAMs-Au substrates, respectively.[25]

The arylamine/anode CT interface is strongly dependent on the chemical nature of the anode as revealed by the work function dependence of $V_t$. We characterize a series of diodes fabricated on Ag, Cu, and Au anodes, respectively. For 4AAPD diodes, $V_t$ increases as the anode work function increases. The lowest $V_t$ is observed in the diode with Ag anode, while the highest $V_t$ is found from the diodes with Au anode. This observation is reasonable with considering that the built-in injection barrier, i.e., the difference between the anode and cathode work function, is increased as the anode work function increases. However, unusual behavior is observed in the CN-4AAPD diodes, where the CN-4AAPD/anode CT interface plays a critical role. Now $V_t$ is not only affected by the built-in barrier but also by the CT interaction at the anode/CN-4AAPD interface. In this case, the CN-4AAPD diode on Cu anode has the highest $V_t$, while the lowest $V_t$ is observed from the diode with Au anode, suggesting that the CT interaction between CN-4AAPD and Au is weaker than that of Ag and Cu.[24] Again it shows that the diode $V_t$ is a powerful tool in studying the organic/metal CT interfaces.

The surface morphology of arylamine molecular films plays an important role in influencing the threshold thickness of the modification interlayer, i.e., by which the CT interface is largely enhanced or impeded, respectively. Figure 5 gives the typical tapping mode AFM images of the 50-nm-thick arylamine films. Both 4AAPD and CN-4AAPD films are composed of small island-shaped domains. As evidenced by the roughness analysis, the films deposited on SAMs-Au are little smoother than that on bare Au. While the CN-4AAPD films are a little rougher than that of 4AAPD. Therefore, when CN-4AAPD is used as the anode modification interlayer, it will need a film thicker than that of 4AAPD so as to form a continue pinhole-free layer to largely induce the CT interface (As confirmed by Fig. 3(c)). This result is also in good agreement with our previous observation, which indicated that 4AAPD could form very smooth film even when the film thickness just reaches about several nanometers.[20]

The conduction mechanism of our diodes is injection limited, which could be well characterized by Richardson-Schottky (RS) thermonic emission model.[1,26] In RS equation, $J=AT^2\exp(-(\Phi_B-\beta E_{eff}^{1/2})/(k_BT))$. Where, J is the current density, A is the Richardson constant, β is a constant with value inversely proportional to the square root of the molecular permittivity; T is the temperature; $\Phi_B$ is the zero-field injection barrier, and $k_B$ is the Boltzmann's constant. As shown in Figure 6, for the diodes fabricated on both Au and SAMs-Au anodes, the plots of Log (I) versus $E_{eff}^{1/2}$ show a linear behavior in a large range of current and electrical field. From the slope and the y axis intercept, we can extract out the information of β and $\Phi_B$, respectively. The slope equals to $\beta/k_BT$ and the y-intercept equals to Log $(AT^2)$-$\Phi_B/k_BT$. As seen, the diodes fabricated on SAMs-Au have a lower y-intercept (i.e. higher $\Phi_B$) than that fabricated bare gold, suggesting that the SAMs introduce an extra injection barrier to the devices. The slope of 4AAPD diodes is higher than that of CN-4AAPD devices, which may indicate that 4AAPD has a lower molecular permittivity than that of CN-4AAPD.

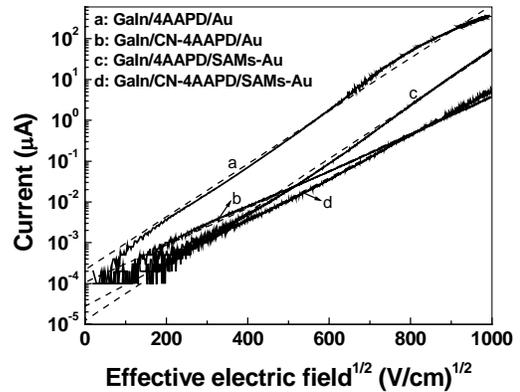

Fig. 6 Semilog plot of current versus the square root of effective electrical field for the 50-nm 4AAPD and CN-4AAPD diodes fabricated on Au and SAMs-Au, respectively.

Based on the chemical nature of the molecular adsorption, organic/metal interfaces could be classified into physisorption and chemisorption, respectively.[9] The presence of a physisorption species would lead to the decrease of the metal work function. On the other hand, two situations are possible for chemisorptions according to the charge transfer degree from the metal to the organic adsorbates: 1) for large charge transfer, the adsorption of such molecules would result in the increase of metal work function. If such molecules are adsorbed on the anode of an organic diode, it will lead to the decrease of the hole injection barrier and thus lower $V_t$. 2) For small charge transfer, the metal work function will be decreased by the adsorbates. In this case, the adsorption of the molecules on the anode will result in diode $V_t$ increasing. This is consistent with our observations for the case of CN-4AAPD/Au interface.

**Conclusion**

In summary, we clearly demonstrate that the arylamine/metal CT interfaces can be characterized through analyzing the I-V data of single-layer organic diodes with the aid of anode interfacial engineering approach. The diode turn-on voltage is very sensitive to the presence of arylamine/metal CT interfaces, which can thus serve as a powerful probe in deterring such interfaces. The arylamine terminal functional groups have critical influence on the CT interfaces with CN-4AAPD and 4AAPD acting as CT enhancing and impeding layers, respectively. The CT interface can be tuned through engineering the chemical nature and thickness of the anode modification layer. Moreover, we show that there exists CT interaction between CN-4AAPD and the surface of Au pre-modified by 1-Decanethiol SAMs, while no resemble behavior could be observed for 4AAPD. The conduction of our diodes is injection limited, which could be well described by Richardson-Schottky thermonic emission model. We believe that our simple approach should be useful in studying the organic/metal CT interfaces of other kinds of organic semiconductors in more real device situations.

The authors thank R. Duncan and K.-Y. Kim for providing the organic materials. We acknowledge The National Science Foundation for supporting this work through the Materials and Research Science and Engineering Center (grants #DMR-98-09423 and #DMR-02-13985).